\documentclass[traditabstract,printers]{aa}

\usepackage{txfonts}

\usepackage{natbib}
\usepackage{graphicx}

\begin{document}
\title{WASP-34b: a near-grazing transiting sub-Jupiter-mass
exoplanet in a hierarchical triple system\thanks{
Radial velocity and photometric data are only available in electronic form at the
CDS via anonymous ftp to\newline
{\tt cdsarc.u-strasbg.fr (130.79.128.5)} or via\newline
http://cdsarc.u-strasbg.fr/viz-bin/qcat?J/A+A/???/A??}}
\author
{B.~Smalley\inst{1},
D.R.~Anderson\inst{1},
A.~Collier Cameron\inst{2},
C.~Hellier\inst{1},
M.~Lendl\inst{3},
P.~F.~L.~Maxted\inst{1},
D.~Queloz\inst{3},
A.~H.~M.~J.~Triaud\inst{3},
R.~G.~West\inst{4},
S.~J.~Bentley\inst{1},
B.~Enoch\inst{2},
M.~Gillon\inst{5,3},
T.~A.~Lister\inst{6},
F.~Pepe\inst{3},
D.~Pollacco\inst{7},
D.~Segransan\inst{3},
A.~M.~S.~Smith\inst{1},
J.~Southworth\inst{1},
S.~Udry\inst{3},
P.~J.~Wheatley\inst{8},
P.~L.~Wood\inst{1},
J.~Bento\inst{8}
}

\authorrunning{B. Smalley et al.}
\titlerunning{WASP-34b}

\institute{
        Astrophysics Group, Keele University,
        Staffordshire, ST5 5BG, United Kingdom
\and
SUPA, School of Physics and Astronomy, University of St. Andrews,
North Haugh, Fife, KY16 9SS, UK
\and
Observatoire de Gen\`{e}ve, Universit\'{e} de Gen\`{e}ve, 51
Chemin des Maillettes, 1290 Sauverny, Switzerland
\and
Department of Physics and Astronomy, University of
Leicester, Leicester, LE1 7RH, UK
\and
Institut d'Astrophysique et de G\'{e}ophysique, Universit\'{e}
de Li\`{e}ge, All\'{e}e du 6 Ao\^{u}t, 17, B\^{a}t. B5C, Li\`{e}ge 1, Belgium
\and
Las Cumbres Observatory, 6740 Cortona Dr. Suite 102,
Santa Barbara, CA 93117, USA
\and
Astrophysics Research Centre, School of Mathematics
\& Physics, Queen's University, University Road, Belfast,
BT7 1NN, UK
\and
Department of Physics, University of Warwick, Coventry CV4 7AL, UK
}

\date{Received <date> / accepted <date>}

\abstract{We report the discovery of WASP-34b, a sub-Jupiter-mass exoplanet
transiting its 10.4-magnitude solar-type host star (1SWASP\,J110135.89-235138.4;
TYC~6636-540-1) every 4.3177~days in a slightly eccentric orbit ($e =
0.038\pm0.012$). We find a planetary mass of $0.59 \pm 0.01$~$M_{\rm Jup}$ and
radius of $1.22 ^{+0.11}_{-0.08}$~$R_{\rm Jup}$. There is a linear trend in the
radial velocities of 55$\pm$4~m\,s$^{-1}$\,y$^{-1}$ indicating the presence of a
long-period third body in the system with a mass $\ga 0.45$~$M_{\rm Jup}$ at a
distance of $\ga1.2$~AU from the host star. This third-body is either a low-mass
star, white dwarf, or another planet. The transit depth ($(R_{\rm P}/R_{*})^{2}$
= 0.0126) and high impact parameter ($b = 0.90$) suggest that this could be the
first known transiting exoplanet expected to undergo grazing transits, but with
a confidence of only $\sim80\%$.}


\maketitle

\section{Introduction}

The majority of the known exoplanets have been discovered using the radial
velocity technique \citep{1995Natur.378..355M}. In recent years, however, an
ever increasing number have been discovered as a result of group-based and
space-based transit search survey projects. Transiting exoplanets allow
parameters such as the mass, radius, and density to be accurately determined, as
well as their atmospheric properties to be studied during their transits and
occultations
\citep{2005ApJ...626..523C,2009MNRAS.394..272S,2009IAUS..253...99W}.

The SuperWASP project has robotic observatories in the Canary Islands and South
Africa. The wide angle survey is designed to find exoplanets around relatively
bright stars in the $V$-magnitude range $9\sim13$. A detailed description of the
SuperWASP project is given in \citet{2006PASP..118.1407P}.

In this paper we report the discovery of WASP-34b, an exoplanet in orbit around
its $V = 10.4$~{mag.} host star 1SWASP\,J110135.89-235138.4 in the constellation
Crater. We present the WASP-South discovery photometry, together with Euler
Telescope photometry and CORALIE radial velocity measurements.

\section{Observations}

\subsection{WASP-South photometry}
\label{WASP_Phot}

\begin{figure}

\includegraphics[height=\columnwidth,angle=-90]{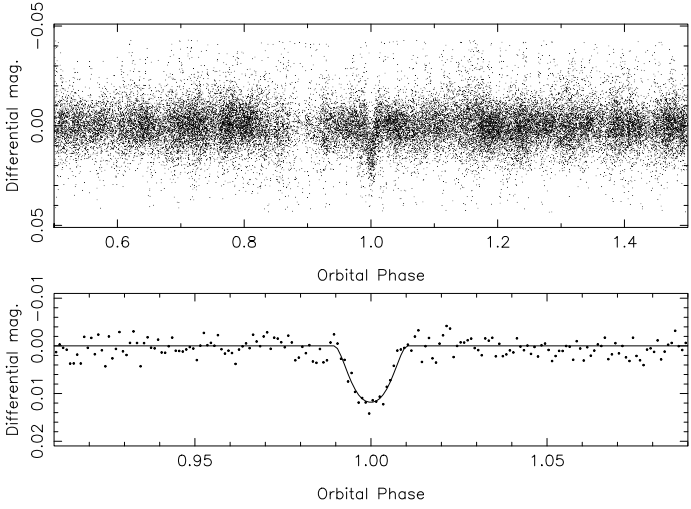}

\caption{WASP photometry of WASP-34 folded on the orbital of period of
4.3177~days (upper panel). The lower-panel shows the photometry binned in 0.001
phase steps and centred on the transit. The solid line is best-fit MCMC solution
(see Sect.~\ref{MCMC})}

\label{WASP-phot}
\end{figure}

The host star WASP-34 (1SWASP\,J110135.89-235138.4; CD-23 9677; TYC 6636-540-1;
GSC 06636-00540) was observed within two WASP-South camera fields during the
periods 2006 May 4 to June 20 and 2007 January 4 to June 1, and in 3 fields
during the period 2008 January 1 to May 28. A total of 35\,351 data points were
obtained. The pipeline-processed data were de-trended and searched for transits
using the methods described in \citet{2006MNRAS.373..799C}, yielding a detection
of a periodic, transit-like signature with a period of 4.3177~days and a depth
of 0.011~magnitudes (Fig.~\ref{WASP-phot}).

\subsection{Spectroscopic observations with CORALIE}
\label{CORALIE_Spec}

Spectroscopic observations were obtained with the CORALIE spectrograph on the
Swiss 1.2m telescope. The data were processed using the standard pipeline
\citep{1996A&AS..119..373B,2000A&A...354...99Q,2002A&A...388..632P}. A total of
24 radial velocity (RV) and line bisector span ($V_{\rm span}$) measurements
were made between 2009 December 1 and August 1 (Table~\ref{rv-data}). The
bisector spans are a measure of the asymmetry of the cross-correlation function
and, based on our experience, have standard errors of $\approx2\sigma_{\rm
RV}$.

\begin{table} 
\caption{Radial velocity (RV) and line bisector spans ($V_{\rm span}$)
measurements for WASP-34 obtained by CORALIE spectra.} 
\label{rv-data} 
\centering
\begin{tabular}{lll} \hline\hline
BJD--2\,400\,000 (UTC)& RV (km\, s$^{-1}$) & $V_{\rm span}$ (km\, s$^{-1}$) \\ \hline
55166.8238 & 49.7903 $\pm$ 0.0044 & $-$0.0413\\
55168.8183 & 49.9372 $\pm$ 0.0043 & $-$0.0295\\
55170.8431 & 49.7923 $\pm$ 0.0042 & $-$0.0275\\
55172.8238 & 49.9253 $\pm$ 0.0046 & $-$0.0540\\
55174.8487 & 49.8141 $\pm$ 0.0041 & $-$0.0499\\
55175.8479 & 49.7973 $\pm$ 0.0039 & $-$0.0376\\
55176.8227 & 49.8806 $\pm$ 0.0042 & $-$0.0327\\
55179.8417 & 49.7888 $\pm$ 0.0041 & $-$0.0350\\
55180.8558 & 49.8611 $\pm$ 0.0041 & $-$0.0401\\
55181.8211 & 49.9414 $\pm$ 0.0042 & $-$0.0382\\
55182.8513 & 49.8765 $\pm$ 0.0049 & $-$0.0405\\
55184.8546 & 49.8432 $\pm$ 0.0044 & $-$0.0282\\
55186.8291 & 49.9058 $\pm$ 0.0046 & $-$0.0142\\
55190.8501 & 49.9152 $\pm$ 0.0045 & $-$0.0438\\
55261.7732 & 49.7686 $\pm$ 0.0049 & $-$0.0303\\
55262.6716 & 49.8191 $\pm$ 0.0041 & $-$0.0375\\
55372.5070 & 49.8731 $\pm$ 0.0050 & $-$0.0401\\
55375.6012 & 49.8797 $\pm$ 0.0070 & $-$0.0128\\
55376.5162 & 49.8956 $\pm$ 0.0080 & $-$0.0249\\
55380.5162 & 49.8922 $\pm$ 0.0048 & $-$0.0383\\
55391.4963 & 49.7631 $\pm$ 0.0053 & $-$0.0275\\
55399.4711 & 49.7695 $\pm$ 0.0048 & $-$0.0249\\
55403.4675 & 49.8159 $\pm$ 0.0049 & $-$0.0133\\
55410.4711 & 49.8913 $\pm$ 0.0049 & $-$0.0437\\
\hline 
\end{tabular} 
\end{table} 

\begin{figure}

\includegraphics[height=\columnwidth,angle=-90]{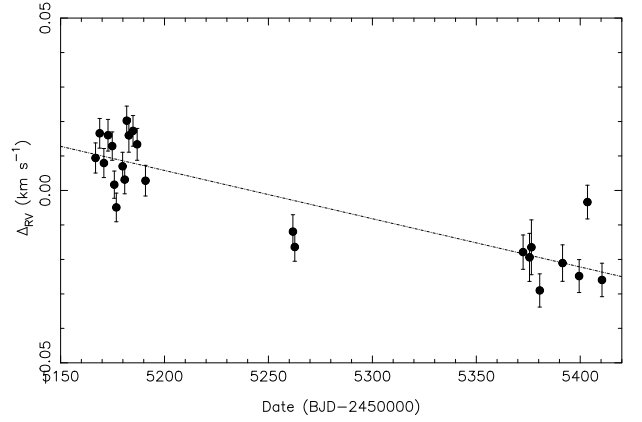}

\caption{Residuals to the periodic orbital fit ($\Delta_{\rm RV}$) as a function
of time. There is a clear long-term trend in the systemic velocity of
55$\pm$4~m\,s$^{-1}$\,y$^{-1}$.}

\label{RV+trend}
\end{figure}

\begin{figure}

\includegraphics[height=\columnwidth,angle=-90]{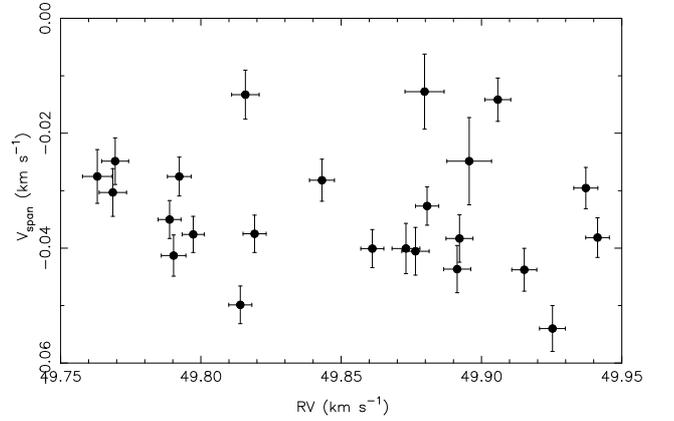}

\caption{Line bisectors ($V_{\rm span}$) as a function of RV for WASP-34.
Bisector uncertainties of twice the RV uncertainties have been adopted.
There is negligible correlation between $V_{\rm span}$ and the stellar RV.}

\label{RV-BS}
\end{figure}

The RV measurements show velocity variations with the same period as the transit
light curve, but with a relatively large scatter in the residuals (16.1
m\,s$^{-1}$) to a circular orbit fit. Analysis of these residuals shows a trend
as a function of time of 55$\pm$4~m\,s$^{-1}$\,y$^{-1}$, with an F-test
revealing that this is significant at $>99.9\%$ level (Fig.~\ref{RV+trend}). The
amplitude of the RV variations and the absence of any correlation with RV of the
line bisector spans ($V_{\rm span}$) in Fig.~\ref{RV-BS} indicates that it is
highly improbable that the RV variations are due to an unresolved eclipsing
binary or chromospheric activity \citep{2001A&A...379..279Q}.

\subsection{Euler Telescope photometry}
\label{Euler_Phot}

We observed two transits (March 3 and April 12, 2010) of WASP-34b with the 1.2m
Euler Telescope. All data were taken in the $R_G$ band and, owing to the
brightness of the target ({mag.} $R$=9.81), with slight defocus. While for the
first transit there was no guiding available at the telescope, we made use of a
new absolute tracking (ABTR) mode for the second observation. The ABTR performs
tracking by computing the location of light sources on each science frame and
matching them with a catalogue. This way, the exact position of each image is
found and drifts from the nominal position are corrected by adjusting the
telescope pointing between the exposures. As a result the target is kept on the
same pixel during the whole transit.

\begin{figure} 

\includegraphics[height=\columnwidth,angle=-90]{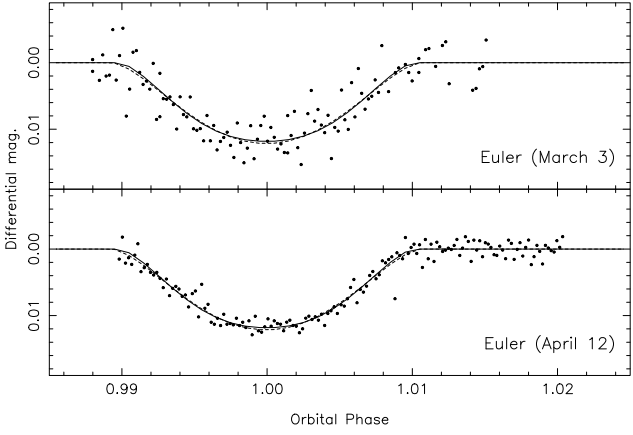}

\caption{Euler Telescope transit photometry. The solid line is the best-fit
main-sequence-constrained MCMC solution, while the dashed-line is that for the
non-main-sequence-constrained solution.}

\label{phot-zoom-euler} 
\end{figure}

The reduction was the same for both transits. After correcting the images for
bias, overscan and flatfield, we performed aperture photometry on the data. The
flux was extracted for all stars in the field and the final light curve was
obtained from differential photometry of the target and a reference source
obtained by combining the 3 brightest reference stars. We reached a r.m.s.
photometric precision of 4.2 {m\,mag.}\,min$^{-1}$ for the transit of March 3
and 1.4{ m\,mag.}\,min$^{-1}$ for the transit of April 12
(Fig.~\ref{phot-zoom-euler}). This improvement can mainly be attributed to the
use of the ABTR. As the stars were kept on the same pixels, flat field effects
were greatly reduced.

\section{Spectral Analysis}

The individual CORALIE spectra of WASP-34 were co-added to produce a single
spectrum with an average S/N of around 100:1. The standard pipeline reduction
products were used in the analysis.

The analysis was performed using the methods given in 
\citet{2009A&A...496..259G} and \citet{2005MSAIS...8..130S}. The H$\alpha$ line
was used to determine the effective temperature ($T_{\rm eff}$), while the Na
{\sc i} D and Mg {\sc i} b lines were used as surface gravity ($\log g$)
diagnostics. The parameters obtained from the analysis are listed in
Table~\ref{wasp34-params}. The elemental abundances were determined from
equivalent width measurements of several unblended lines. A value for
microturbulence ($\xi_{\rm t}$) was determined from Fe~{\sc i} using the method
of \citet{1984A&A...134..189M}. The quoted error estimates include that given by
the uncertainties in $T_{\rm eff}$, $\log g$ and $\xi_{\rm t}$, as well as the
scatter due to measurement and atomic data uncertainties.

The projected stellar rotation velocity ($v \sin i$) was determined by fitting
the profiles of several unblended Fe~{\sc i} lines. A value for macroturbulence
($v_{\rm mac}$) of 2.3 $\pm$ 0.3 km\, s$^{-1}$ was assumed, based on the recent
calibration of \citet{2010MNRAS.405.1907B}, and an instrumental FWHM of 0.11
$\pm$ 0.01 \AA, determined from the telluric lines around 6300\AA. A
best-fitting value of $v \sin i$ = 1.4 $\pm$ 0.6~km\, s$^{-1}$ was obtained.

\begin{table}
\caption{Stellar parameters of WASP-34.}
\centering
\begin{tabular}{ll} \hline\hline
Parameter  & Value \\ \hline
RA (J2000.0)   & 11h01m35.89s \\
Dec (J2000.0)  & $-$23{\degr}51{\arcmin}38.4{\arcsec} \\
$V$ {mag.}     & 10.4 \\
Distance       & $120 \pm 15$~pc \\
$T_{\rm eff}$  & 5700 $\pm$ 100 K \\
$\log g$       & 4.5 $\pm$ 0.1 \\
$\xi_{\rm t}$  & 0.9 $\pm$ 0.1 km\, s$^{-1}$ \\
$v \sin i$     & 1.4 $\pm$ 0.6 km\, s$^{-1}$ \\
{[Fe/H]}       &$-$0.02 $\pm$ 0.10 \\
{[Si/H]}       &  +0.08 $\pm$ 0.08 \\
{[Ca/H]}       &  +0.06 $\pm$ 0.15 \\
{[Ti/H]}       &  +0.09 $\pm$ 0.06 \\
{[Cr/H]}       &$-$0.02 $\pm$ 0.07 \\
{[Ni/H]}       &  +0.03 $\pm$ 0.07 \\
log A(Li)      &$<$0.8  \\
Spectral Type  & G5 \\
$M_{\star}$    & 1.01 $\pm$ 0.07  $M_{\sun}$ \\
$R_{\star}$    & 0.93 $\pm$ 0.12  $R_{\sun}$ \\
\hline
\end{tabular}
\label{wasp34-params}
\tablefoot{The spectral type is that listed as G5 in the Yale Zone Catalogue
\citep{1943TOYal..14....1S}, which  agrees with that implied from $T_{\rm eff}$
\citep[Table~B.1]{2008oasp.book.....G}. $M_{\star}$ and $R_{\star}$ are
estimated using the \citep{2010A&ARv..18...67T} calibration.}
\end{table}

The lack of any measurable lithium in the CORALIE spectra implies that WASP-34
is a relatively old star, with an age of ${\ga}5$\,Gy
\citep{2005A&A...442..615S}. The measured $v \sin i$ of WASP-34 implies a
rotational period of $P_{\rm rot} \simeq 34\pm15$~days, which yields a
gyrochronological age of $\sim6.7^{+6.9}_{-4.5}$~Gy using the relation of
\citet{2007ApJ...669.1167B}, which is consistent with the lithium age. In many
respects WASP-34 is very similar to the Sun. 

\section{Planetary system parameters}
\label{MCMC}

To determine the planetary and orbital parameters the CORALIE radial velocity
measurements were combined with the photometry from the WASP and Euler
Telescopes in a simultaneous fit using the Markov Chain Monte Carlo (MCMC)
technique. The details of this process are described in
\citet{2007MNRAS.380.1230C} and \citet{2008MNRAS.385.1576P}. Four sets of
solutions were used: with and without the main-sequence mass-radius constraint
for both circular and floating eccentricity orbits. In all cases we have
included a linear trend in the systemic velocity, $\frac{d\gamma}{dt}$, as a
free parameter. Limb-darkening uses the four-coefficient model of
\cite{2000A&A...363.1081C} for the $r$-band, for both WASP and Euler photometry.

\begin{table}
\caption{System parameters for WASP-34b.}
\label{wasp34b-mcmc}
\centering
\begin{tabular}{ll} \hline\hline
Parameter & Value \\ \hline
Transit epoch (HJD), $T_0$ &
2454647.55358 $\pm$ 0.00064
\\[+1mm]
Orbital period, $P$  &
4.3176782 $\pm$ 0.0000045  d
\\[+1mm]
Transit duration, $T_{14}$ &
0.0876 $^{+0.0016}_{-0.0017}$ d
\\[+1mm]
Transit depth, $(R_{\rm P}/R_{*})^{2}$ &
0.01261 $\pm$ 0.00026
\\[+1mm]
Impact parameter, $b$ &
0.904 $^{+0.017}_{-0.014}$
\\[+1mm]
Stellar reflex velocity, $K_1$ &
0.0721 $\pm$ 0.0012 km\,s$^{-1}$
\\[+1mm]
Centre-of-mass velocity at time $T_0$, $\gamma$ &
49.9395 $\pm$ 0.0064 km\,s$^{-1}$
\\[+1mm]
Drift in centre-of-mass velocity, 
$\displaystyle \frac{d\gamma}{dt}$ &
$-$54.8 $\pm$ 4.0 m\,s$^{-1}$\,y$^{-1}$
\\[+1mm]
Orbital separation, $a$ &
0.0524 $\pm$ 0.0004 AU
\\[+1mm]
Orbital inclination, $i$ &
85.2 $\pm$ 0.2 {\degr}
\\[+1mm]
Orbital eccentricity, $e$ &
0.038 $\pm$ 0.012
\\[+1mm]
Argument of periastron, $\omega$ &
$-$40.2 $^{+23.2}_{-18.6}$ {\degr}
\\[+1mm]
\hspace{3cm} $e \cos(\omega)$ &
$+$0.028 $\pm$ 0.009
\\[+1mm]
\hspace{3cm} $e \sin(\omega)$ &
$-$0.024 $\pm$ 0.016
\\[+1mm]
Planet mass, $M_{\rm P}$ &
0.59 $\pm$ 0.01 $M_{\rm Jup}$
\\[+1mm]
Planet radius, $R_{\rm P}$ &
1.22 $^{+0.11}_{-0.08}$ $R_{\rm Jup}$
\\[+1mm]
Planet surface gravity, $\log g_{\rm P}$ &
2.96 $^{+0.05}_{-0.07}$
\\[+1mm]
Planet density, $\rho_{\rm P}$ &
0.34 $\pm$ 0.08 $\rho_{\rm Jup}$
\\[+1mm]
Planet equilibrium temperature, $T_{\rm P}$ &
1250 $\pm$ 30 K
\\[+1mm]
\hline
\end{tabular}
\tablefoot{The planet equilibrium temperature, $T_{\rm P}$, assumes a Bond
albedo of $A=0$ and even redistribution of heat around the planet.}
\end{table}

\begin{figure}

\includegraphics[height=\columnwidth,angle=-90]{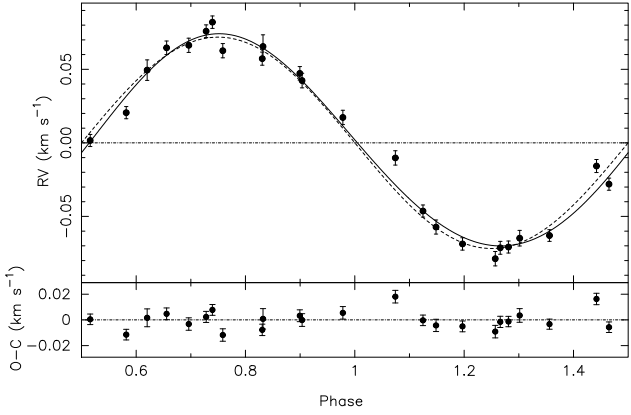}

\caption{Radial velocity measurements relative to the systemic-mass velocity
including the long-term drift, $\frac{d\gamma}{dt}$, compared to the eccentric
orbital solution for WASP-34b (solid line). The dashed-line is the corresponding
circular orbital solution.}

\label{RV+resid}
\end{figure}

With the main-sequence constraint imposed but the eccentricity floating, a
value  of $e = 0.038 \pm 0.012$ is found, which is significant at the 99.3\%
level \citep[Eq. 21]{1971AJ.....76..544L}. The radial velocity measurements are
well fitted by this orbital solution and the linear long-term drift
(Fig.~\ref{RV+resid}). However, a F-test is a more reliable guide to the
significance of the eccentricity \citep[Eq. 27]{1971AJ.....76..544L}. The
eccentric orbit fit has $\chi^2 = 57.4$, compared to $\chi^2 = 68.5$ for a
circular orbit.  We adopt the null hypothesis that the underlying orbit is truly
circular but recognise that an eccentric fit to noisy and unevenly sampled data
will always lead to a spurious improvement in $\chi^2$. The probability that an
improvement in the fit could have arisen by chance with the addition of 2
fitting parameters, $e \cos(\omega)$ and $e \sin(\omega)$, is 17.1\%. While this
is too high to confidently claim detection of eccentricity, it does not confirm
that the orbit is truly circular. Hence, given there is some evidence for a
non-circular orbit, we present the system parameters for the eccentric solution
(Table~\ref{wasp34b-mcmc}). The impact parameter and transit depth suggest that
the planet could be in a grazing transit.

Relaxing the main-sequence constraint yields a larger stellar radius
(1.20$\pm$0.09~$R_{\sun}$) and lower surface gravity ($\log g = 4.3$), which is
marginally inconsistent with the spectroscopic $\log g$. The planetary radius
increases to 1.82$^{+0.82}_{-0.48}$~$R_{\rm Jup}$ and the impact parameter also
rises to $0.98^{+0.08}_{-0.06}$. The planetary density now drops to
$0.11^{+0.15}_{-0.08}$~$\rho_{\rm Jup}$, which would be similar to that found
for WASP-17b \citep{2010ApJ...709..159A}. While the fit to the transit
lightcurves is only slightly different (Fig.~\ref{phot-zoom-euler}), the
uncertainties in the planetary parameters are now rather large, indicating that
the near-grazing nature of the transit is poorly constrained by the available
photometry. Consequently, we adopt the main-sequence-constrained eccentric
solution.

\section{Discussion}

WASP-34b is a sub-Jupiter mass exoplanet transiting a G5 host star every 4.3177
days. A simultaneous fit to transit photometry and radial-velocity measurements
gave a planetary mass of $0.59 \pm 0.01$~$M_{\rm Jup}$ and radius of $1.22
^{+0.11}_{-0.08}$~$R_{\rm Jup}$.

In many respects the WASP-34b system looks like transiting version of 51 Peg
\citep{1995Natur.378..355M} with similar periods, separations, masses and host
stars. Of the known transiting system, WASP-34b joins an increasing number of
exoplanets with masses $\sim0.5M_{\rm Jup}$ and radii $\sim1.0R_{\rm Jup}$, such
as WASP-22b \citep{2010AJ....140.2007M}, WASP-25b \citep{2010arXiv1009.5917E},
HAT-P-24b \citep{2010arXiv1008.3389K} and HAT-P-25b \citep{2010arXiv1008.3565Q}.
Of these systems, the similarity to WASP-22b is striking considering that it too
has a linear trend in radial velocities due to a third-body in the system.

\subsection{Properties of the third body}

The significant linear trend in the radial velocities of 55
$\pm$4~m\,s$^{-1}$\,y$^{-1}$, noted in Sect.~\ref{CORALIE_Spec}, indicates the
presence of a third body in the system with a long period. Using the constant
acceleration method of \cite{2009ApJ...703L..99W} the third body satisfies
\begin{equation}
\frac{M_c \sin i_c}{{a_c}^2} = 0.307 \pm 0.022,
\end{equation}
where $M_c$ and $a_c$ are the mass (in $M_{\rm Jup}$) and orbital separation (in
AU) of the third body. Given that the period of this outer body must be greater
than at least twice the RV data baseline, $P_c \ga 490$~days and, using Kepler's
third law, $a_c \ga 1.2$~AU. Hence $M_c \ga 0.45$~$M_{\rm Jup}$.

Using literature photometry we estimated the bolometric flux at the Earth to be
$f_{\rm bol} = 1.97 \pm 0.10 \times 10^{-9}$~erg\,s$^{-1}$\,cm$^{-2}$, which
gives $T_{\rm eff}$ =  5740 $\pm$ 140~K using the Infrared Flux Method (IRFM)
\citep{1977MNRAS.180..177B}, which is in excellent agreement with the
spectroscopic temperature determination. The modified IRFM method of
\cite{1993MNRAS.265.1035S} suggests that a main-sequence companion would be
cooler than $\sim$4000~K (M-type or later). There is a {\it GALEX}
\citep{2007ApJS..173..682M} near-UV flux measurement for WASP-34, and this is in
in agreement with that predicted for a star of this temperature. This is no
significant UV excess that could be attributed to a hot compact stellar
companion, but this does not exclude the possibility of a cool white dwarf
companion. Thus, we expect that the companion object is either a low mass star
(M-type or cooler), cool white dwarf or another planet in a wide long-period
orbit. Further radial velocity measurements are required to constrain the orbit.

\subsection{Grazing nature of the system}

\begin{figure}

\includegraphics[height=\columnwidth,angle=-90]{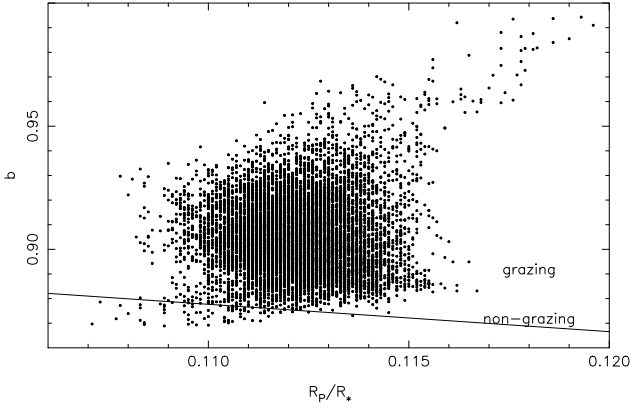}

\caption{Plot of the MCMC posteriors $b$ and ${R_{\rm P}}/{R_{\star}}$. The
solid-line indicates the position of the stellar limb, i.e. $b+{R_{\rm
P}}/{R_{\star}} = 1$. A total of 89.6\% of the points lie above the line and are
grazing solutions.}

\label{grazing}
\end{figure}

The transit depth ($(R_{\rm P}/R_{*})^{2}$ = 0.0126) and high impact parameter
($b = 0.90$) suggest that WASP-34b could be the first transiting exoplanet known
to undergo grazing transits. For a transit to be truly grazing, we require
\begin{equation}
b+{R_{\rm P}}/{R_{\star}} > 1
\end{equation}
and that this {\it grazing criterion} is significant compared to the
observational uncertainties. For WASP-34b we find a value of 1.016
$^{+0.017}_{-0.014}$ for the {\it grazing criterion}. Hence, we might expect
grazing transits, unlike the near-grazing transits of TrES-2
\citep{2006ApJ...651L..61O} and HAT-P-14b \citep{2010ApJ...715..458T}, which
have {\it grazing criteria} of 0.972$\pm$0.007 and 0.968$\pm$0.022,
respectively. Figure~\ref{grazing} shows the 10\,000 MCMC posteriors for
${R_{\rm P}}/{R_{\star}}$ and $b$. A total of 89.6\% of these satisfy the {\it
grazing criterion}. If the {\it grazing criterion} were equal to unity, then we
would expect 50\% of the points to be higher and 50\% to be lower than one.
Using the odds ratio test \citep{2010arXiv1008.3389K}, we find a 82.8\%
probability (1.4-$\sigma$) that the system is truly grazing. However, this is
lower than a 3-$\sigma$ value which would be a reasonable limit for detection of
a truly grazing system. Hence, we conclude that WASP-34b has near-grazing
transits.

The eccentricity ($e$) and argument of periastron ($\omega$) are such that the
occultation impact parameter will be \begin{equation}
b_{\rm occ} = (0.953\pm0.031) \, b_{\rm
tra} = 0.862\pm0.035.
\end{equation}  Hence, $b_{\rm occ}+{R_{\rm P}}/{R_{\star}} =
0.974\pm0.035$ with only 19.2\% of the MCMC posteriors greater than 1.
Therefore, the occultations of the planet behind the star are expected to be
total. However, the uncertainties in the orbital elements are such that is a
possibility that the occultations could be partial.

For transits with large $b$, the limb-darkening coefficients are extremely
correlated to both $R_{\rm P}/R_{*}$ and $b$ \citep{2010arXiv1006.5680K}. In
order to investigate the effects of limb-darkening on the inferred grazing
nature of the system, we re-ran our MCMC analysis using limb-darkening
coefficients for different filter bands (Table~\ref{limb}). Compared to the
$r$-band results, there are small but noticeable, differences. Hence, this
heuristic demonstrates that there is a $75\sim85$\% probability that the transit
is actually grazing.

\begin{table}

\caption{Effect of limb-darkening on the derived impact parameter, $b$, and
transit depth, $(R_{\rm P}/R_{*})^{2}$, for various filter bands. The grazing
criterion, $b+{R_{\rm P}}/{R_{\star}}$, is given along with the percentage of
MCMC posteriors that are grazing,\%($>$1), and the confidence values, P($>$1).}
\centering

\begin{tabular}{llllll}\hline\hline
band & $b$ & $(R_{\rm P}/R_{*})^{2}$ & $b+{R_{\rm P}}/{R_{\star}}$  &\%($>$1) & P($>$1)\\ \hline
$v$ & 0.898  &   0.01309   &    1.012$^{+0.017}_{-0.013}$  &  83.1 & 74.7 \\[+1mm]
$r$ & 0.904  &   0.01261   &    1.016$^{+0.017}_{-0.014}$  &  89.6 & 82.8 \\[+1mm]
$i$ & 0.905  &   0.01213   &    1.015$^{+0.016}_{-0.012}$  &  91.1 & 84.9 \\[+1mm]
$z$ & 0.907  &   0.01190   &    1.016$^{+0.016}_{-0.013}$  &  90.3 & 83.8 \\[+1mm]
\hline
\end{tabular}
\label{limb}
\end{table}

The near-grazing nature of the system makes it sensitive to additional planets,
via changes to transit duration and shape. Given that there is evidence for
another object in the WASP-34 system, further monitoring of the transits is
required.

\section*{Acknowledgments}

WASP-South is hosted by the South African Astronomical Observatory and their
support and assistance is gratefully acknowledged. M. Gillon acknowledges
support from the Belgian Science Policy Office in the form of a Return Grant. 
We thank the referee, David Kipping, for his constructive and helpful comments
on the original manuscript.

\bibliographystyle{aa}

\end{document}